\documentclass[conference]{IEEEtran}
\IEEEoverridecommandlockouts

\usepackage{cite}
\usepackage{amsmath,amssymb,amsfonts}
\usepackage{algorithmic}
\usepackage{graphicx}
\usepackage{textcomp}
\usepackage{xcolor}
\def\BibTeX{{\rm B\kern-.05em{\sc i\kern-.025em b}\kern-.08em
    T\kern-.1667em\lower.7ex\hbox{E}\kern-.125emX}}

\usepackage{float}    
\usepackage{subfig}

\usepackage{tcolorbox}
\usepackage{todonotes}

\author{\IEEEauthorblockN{Ikram Messadi, Giulia Cervia, Vincent Itier}
\IEEEauthorblockA{IMT Nord Europe, Institut Mines-Télécom, Centre for Digital Systems, F-59000 Lille, France\\
Emails: \{ikram.messadi, giulia.cervia,  vincent.itier\}@imt-nord-europe.fr}
}

\begin{document}

\title{Image selective encryption analysis using mutual information in CNN based embedding space}

\maketitle

\begin{abstract}
As digital data transmission continues to scale, concerns about privacy grow increasingly urgent —yet privacy remains a socially constructed and ambiguously defined concept, lacking a universally accepted quantitative measure. This work examines information leakage in image data, a domain where information-theoretic guarantees are still underexplored.
At the intersection of deep learning, information theory, and cryptography, we investigate the use of mutual information (MI) estimators — in particular, the empirical estimator and the MINE framework — to detect leakage from selectively encrypted images. 
Motivated by the intuition that a robust estimator would require a probabilistic frameworks that can capture spatial dependencies and residual structures — even within encrypted representations - our work represent a promising direction for image information leakage estimation.

\end{abstract}


\section{Introduction}
Images are among the most common forms of data shared online, and with the widespread use of cloud storage, users frequently upload images to the web. Regardless of content sensitivity, image privacy remains a critical concern.

This issue extends to machine learning and deep learning applications involving visual data. As these models become more powerful, industries such as healthcare, finance, and law increasingly rely on them for automated, robust, and efficient decision-making. However, when models are trained on sensitive data, there is a risk of \textit{information leakage}~\cite{liu2020privacy}.

In response to this risk, recent research has focused on privacy-preserving deep learning, particularly for sensitive visual data like medical images~\cite{huang2022privacy}. Given the importance of this topic, various methods have been proposed to assess image privacy, such as in~\cite{ahmad2010efficiency}.

While Differential Privacy~\cite{dwork2006differential,dwork2008differential, dwork2014algorithmic} has become the gold standard for protecting individual data through randomized mechanisms, more recently, information-theoretic approaches have been proposed to quantify privacy and data leakage in machine learning systems~\cite{bloch_overview_2021}. In this context, \emph{Mutual Information} (MI) emerges as a natural candidate for privacy quantification, given its foundational role in communication theory, its operational meaning, and its connections to cryptography~\cite{wang2016relation}.

Despite its promise, little work has explored mutual information specifically for encrypted image data. Motivated by the long-term goal of integrating privacy-preserving and cryptographic techniques, we build on the approach of~\cite{kim2024crypto} to provide an initial attempt at quantifying image leakage in cryptographic systems using mutual information.

To provide dynamic access control and protect against progressive leakage, we explore selective encryption schemes~\cite{puteaux2021combining} and evaluate different mutual information metrics between original and encrypted images.

Traditionally, mutual information for images is estimated empirically via histograms, which neglect spatial relationships in higher-dimensional data~\cite{larkin2016reflections}. Our work addresses this gap by investigating mutual information in the context of images—arguably one of the first to do so.

In fact, since empirical methods often fail to capture spatial structure, we implement the MINE framework~\cite{belghazi2018mine} for estimating mutual information in images. Furthermore, considering the proven effectiveness of convolutional neural networks (CNNs) in tasks like object detection and segmentation~\cite{sun2022survey, ding2024edge} or image quality assessment~\cite{zhang2018unreasonable}, we enhance the MINE estimation by working on the CNN's embedding of images.




\begin{figure}[t]
    \centering
    \includegraphics[width=1\linewidth]{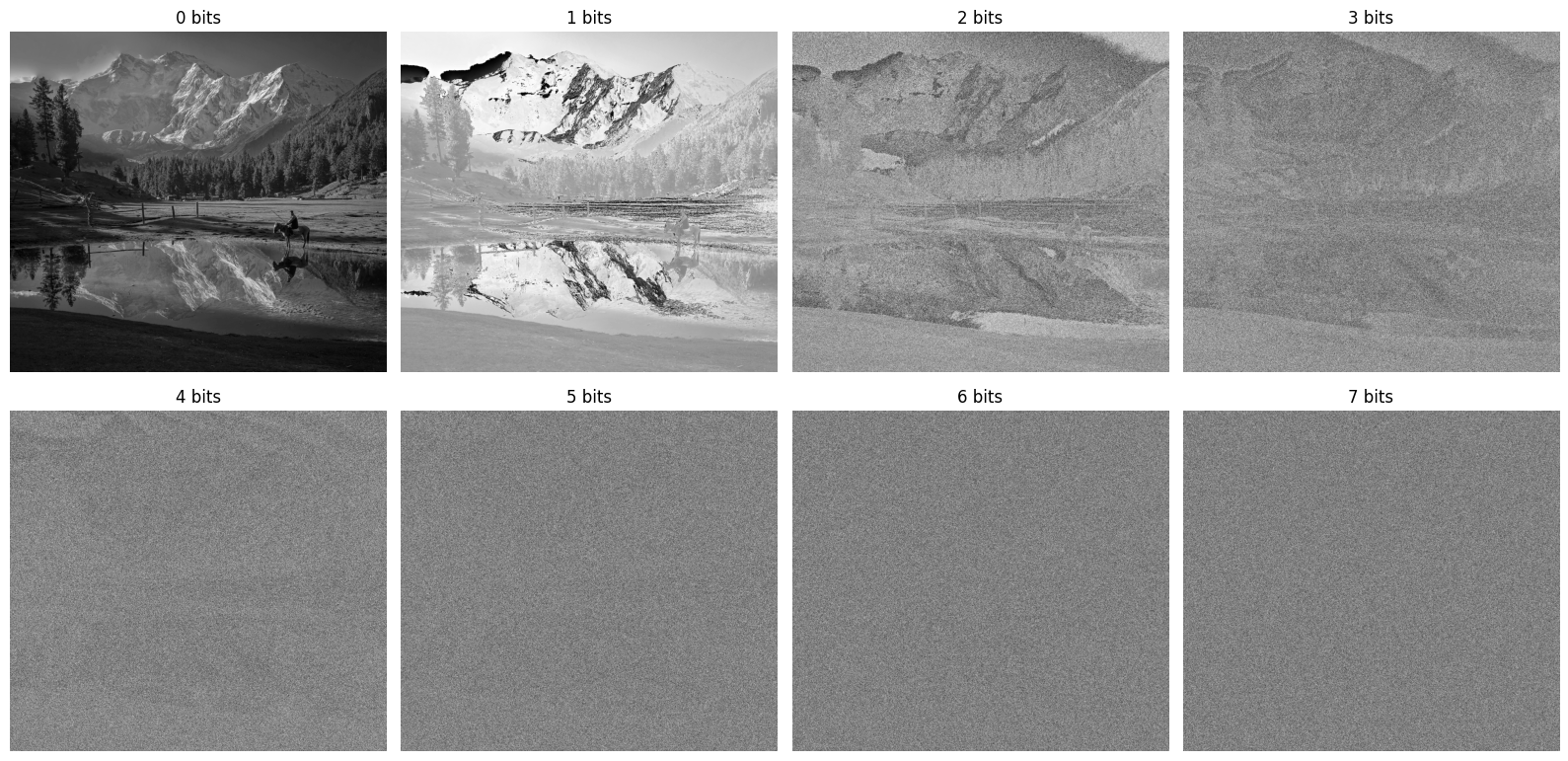}
    \caption{Progressive encryption of pixels according to the number of encrypted $s$ leading bits for an 8-bits encoded grayscale image.}
    \label{fig:selective-encryption}
\end{figure}

\section{Image selective encryption}
\label{section:ise}
Information leakage can occur when an encryption scheme leaves portions of data unencrypted — a method known as selective encryption. Its purpose varies depending on which parts of the data are encrypted and to what extent. For example, one can apply selective encryption with the expectation that the entire stream will be unusable without decrypting the protected portion~\cite{lookabaugh2004selective}. On the contrary, selective encryption can be used to mask sensitive content while allowing some processing. For instance, in~\cite{puteaux2021combining}, authors apply selective encryption on images in order to visually mask the content while allowing integrity checking \emph{i.e.} to verify whether the image has been manipulated or not. 

In this paper, as we want to investigate information leakage, we follow the use case proposed in~\cite{puteaux2021combining}. We note that, since  many different selective encryption methods exist, we believe our analysis can be extended to other approaches and it would be an interesting line of future research.

In our chosen approach, varying levels of visual masking for confidentiality are achieved by encrypting the first significant $s$ bits of each pixel in an 8-bit encoded image, as illustrated in Figure~\ref{fig:selective-encryption}.
The method encrypts only the \textit{Most Significant Bits} (MBS), \emph{i.e.} the bits whose corresponding pixel’s binary representation has the greatest impact on its visual appearance (low frequencies) and perceived brightness. More precisely, for a $m \times n$ grayscale 8-bits image, a pixel is defined as:
\begin{equation}\label{eq: pixel}
    p(i,j) = \sum_{k=0}^{7}p^k(i,j) \times 2^k \text{ with } 0 \leq i\leq m \text{ and } 0 \leq j \leq n, 
\end{equation}
where $p^k(i,j)$ is the bit of the pixel $p(i,j)$ at index $k$.
Selective encryption is applied by using a bitwise {\sc xor} between pixels binary representation and a pseudo-random uniform generated sequence, denoted $b$, of size $s \times n \times m$, such as:
\begin{equation}
    p^{k}_{\textsc{e}}(i,j) = p^k(i,j) \oplus b^k(i,j).
\end{equation}


Moreover, according to Kerckhoff's principle, one can assume that the number of encrypted bits is known.
 Therefore, in this work, the information leakage is investigating on the clear portion $p_{\textsc{c}}(i,j)$ of the pixel $p_{\textsc{e}}(i,j)$ \emph{i.e.} by bit-shifting encrypted pixels: 
\begin{equation}\label{eq: sel_enc}
    p_{\textsc{c}}  = (p_{\textsc{e}} \ll s) \gg s. 
\end{equation}
Furthermore, although we have restricted our explanation to black and white pictures, the extension to color images is straightforward. 


\section{Mutual Information Estimation}

Mutual information's  role in  cryptography dates back to Shannon's cipher system, and for two random variables $X$ and $Y$ it is defined as 
\begin{align}
    I(X;Y)
    &= H(Y) - H(Y|X)\nonumber\\
    &=\mathbb D_{KL}(P_{XY} \parallel P_X P_Y ), \label{eq:mi_kl}
\end{align}where the latter is the Kullback-Leiber divergence.

While mutual information (MI) is a natural measure of correlation between random variables, its practical use depends heavily on accurately estimating the joint distribution \( P_{XY} \) and its marginals. In the absence of better tools, histograms are typically used to approximate the joint distribution.

This useful yet imprecise tool becomes  even less reliable when applied to the context of pictures.
When target variables are a pair composed of an image and its encrypted version, this leads to a natural question about the applicability of MI as a scalable measure in higher dimensions. The answer is not trivial, since the histogram does not take correlation between neighbouring pixels into account.

A notable work, referred to as MINE~\cite{belghazi2018mine}, has proposed to estimate MI in high dimensions, where traditional methods struggle, through leveraging neural networks to approximate  MI efficiently.

The neural network approximates  MI by parameterizing the variational lower bound of the KL-divergence with a neural network. Specifically, it maximizes the Donsker-Varadhan representation of the KL divergence~\cite{belghazi2018mine} which is given by the following: 
\begin{equation}\label{eq:donsker-v}
    \mathbb D_{KL}(\mathbb{P} \parallel \mathbb{Q}) = \sup_{T \in \mathcal{F} }\left[\mathbb{E}_{\mathbb{P}}[T] - \mathbb{E}_{\mathbb{Q}}[\\e^{T - 1}]\right],
\end{equation}where $\mathcal{F}$ is any class of functions. Hence, from \eqref{eq:mi_kl} and \eqref{eq:donsker-v}, the parametrized MI can approximated as: 
\begin{equation}\label{eq: param_mi}
  I_\Theta(X, Y) = \sup_{\theta \in \Theta} \left[\mathbb{E}_{\mathbb{P}_{X Y}}[T_\theta] 
- \log \left(\mathbb{E}_{\mathbb{P}_X \otimes \mathbb{P}_Y}[e^{T_\theta}]\right)\right],  
\end{equation}where $T_{\theta}$ is a parametrized neural network known as the \textit{T statistic network}.
The formulation in \eqref{eq:donsker-v} turns KL divergence estimation into an \textit{optimization problem} across all possible functions $T$.

The MINE framework has proved to be a valuable tool for crypto-analysis in the context of 1D variables~\cite{kim2024crypto}. 
It has not, however,  been extensively studied for the 2D picture problem.
In this context, the challenge is to derive a leakage estimation that can capture any pattern or structure in the image using MINE.

\section{Main contribution}
\label{section:mc}
Building on the challenges discussed previously, this work evaluates information leakage in selectively encrypted images. 

Before analyzing more sophisticated metrics, we first experimented with empirical mutual information to assess leakage between a plain image and its encrypted version.
We start by observing that with the  method presented in Eq.~\eqref{eq: sel_enc}, information theory gives us the following upper bound on the mutual information between the original image $X$ and its encrypted version $Y$:
\begin{align}
    I(X;Y) & \overset{(a)}{=} H(f(X))- H(f(X)|X)\nonumber\\
    &\overset{(b)}{=} H(f(X)) \overset{(c)}{\leq} \log \lvert f(\mathcal X) \rvert = \log \lvert \mathcal Y \rvert \label{eq: mi_Y},
\end{align}where $(a)$   follows from the fact that the encryption is a \textit{deterministic} function $f:  X \mapsto Y$. Further, $(b)$ follows from the definition of relative entropy and the deterministic nature of $f$, that is $ H(f(X)|X)$ is equal to
\begin{align}
- \sum_{x \in \mathcal X} \sum_{f(x) \in f(\mathcal X)} P_{(X, f(X))}(x,f(x)) \underbrace{\log \underbrace{{P_{f(X)|X}(f(x)|x)}}_{=1}}_{=0},
\end{align}and $(c)$ comes from the properties of entropy, and it is an equality for uniform random distributions. 

The upper bound in Eq.~\eqref{eq: mi_Y} represents a worst case scenario, since intrinsically pictures have structure, and their distribution should be very different from the uniform one. However, we observe that by calculating the empirical mutual information on pictures, we observe the same behavior as if we were in the random uniform case. 

Note that, if the pixel were to be uniformly distributed, Eq.~\eqref{eq: mi_Y} would take the simpler formulation of \( I(X;Y)~=~\log\lvert \mathcal Y \rvert\). Then, by combining it with~\eqref{eq: pixel} and~\eqref{eq: sel_enc}, it is easy to see that the mutual information in this case scales with the number of selectively encrypted pixels $s$. 

\begin{figure}[h!]
    \begin{center}
    \includegraphics[width=0.75\columnwidth]{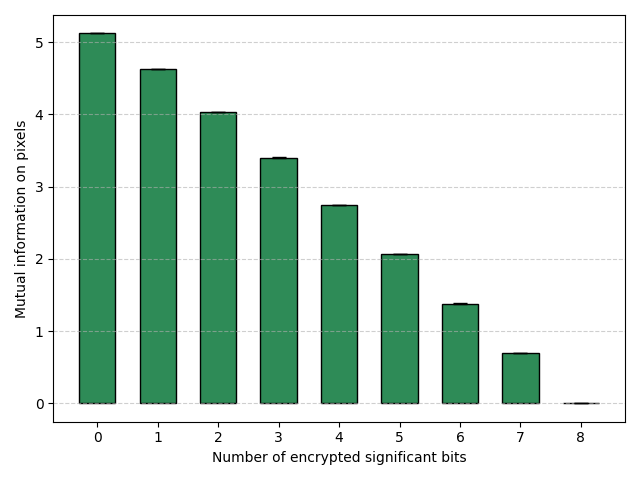}
    \caption{On 100 images from the COCO dataset~\cite{lin2014microsoft}, average scores in terms of Mutual Information estimated on pixels histogram, as a function of the number of encrypted bits $s$.}
    \label{sub:mipix}
    \end{center}
\end{figure}

While theoretically this result would only serve as a bound for real images, Fig.~\ref{sub:mipix} shows that empirical mutual information of real pictures scales linearly with the number of encrypted pixels \( s \), mirroring what one would expect if, instead of a database of images, the pixels were uniformly distributed.

Motivated by this observation, we seek a different mutual information estimator whose behavior with respect to \( s \) deviates from that observed with uniformly distributed random pixels, by accounting for the structure inherent in the image itself. In particular, we are interested in estimators whose response is \emph{not linear} with respect to \( s \).

We emphasize that our work is interested in 
comparing the behavior of different MI estimators — such as the empirical method and MINE — rather than to evaluate the absolute value of mutual information. While MI values naturally vary across different estimators, our focus was on how they respond to increasing numbers of encrypted bits. 

Specifically, we sought an MI estimator whose values decrease \emph{consistently but not linearly} as more bits are encrypted, reflecting the progressive loss of structure illustrated in Fig.~\ref{fig:selective-encryption}.

As anticipated, the first candidate in this research has been the MINE algorithm, and more specifically, our investigation has involved the following two steps: 

\begin{itemize}
    \item  We examine how the existing deep-learning-based MI estimator, MINE~\cite{belghazi2018mine}, performs when applied to images encrypted using the selective encryption approach of~\cite{puteaux2021combining}.
    \item We propose a convolution-based approach to better capture the inherent spatial dependencies in high-dimensional image data to some level, thereby mitigating MINE’s  structural limitations.
\end{itemize}
We also note that the proposed approaches target partially encrypted images, since we consider in this paper that confidentiality is linked to determining the content of the image.

\subsection{MINE on Selectively Encrypted Images}
\label{sec:mine-analysis}
To further question the leakage measure for images, first we have started by reproducing the Mutual Information Neural Estimator model (MINE) as described in~\cite{belghazi2018mine}.
According to our needs, we have used a simplified neural network for the estimation of the function $T$, as introduced in Eq.~\eqref{eq: param_mi}. In our implementation, both the original image and its selectively encrypted version are treated as one-dimensional arrays, an operation we sometimes refer to as \emph{flattening}.

As shown in Section~\ref{sec : exp}, MINE's behavior is not linear with respect to the number of encrypted bits, making it a more appealing metric than empirical MI. However, flattening the pixel into one array dismantles some of the structure of their joint distribution, which leads us to propose an enhanced version of the method, as  explained in the next section.


\subsection{Convolution-Based Mutual Information Estimation}
\label{sec:conv-mi}
Recognizing that existing information-theoretic approaches do not explicitly model spatial dependencies, we explored a convolution-based framework for MI estimation tailored to image data. 
Given the demonstrated success of Convolutional Neural Networks (CNN) in capturing spatial structure~\cite{younesi2024comprehensive,deneu2021convolutional}, the proposed approach divides the image into patches under the premise that critical information is often localized within specific regions. By learning to capture the relationships among neighboring patches, this method aims to incorporate local spatial dependencies into the MI estimation process by encoding the relationship into several feature maps. 
Although it does not fully address the overall pixel distribution, it provides an avenue to address limitations of traditional empirical MI estimation.
We shall also note that convolution in this context is used as a pre-processing framework for the MINE estimator.

We recall that a convolutional neural network (CNN)~\cite{krizhevsky2012imagenet}  is an Artificial Neural Network architecture that uses the convolution operator in one of its layers. That is, convolution combines two functions to produce a third, capturing how the shape of one is influenced or altered by the other.
Commonly, this is used as an encoding of some input space (\emph{e.g.} images), with regularized linear activations
and uses several layers of the same encoding (\emph{i.e.} convolutional filters) and pooling layers (\emph{e.g.} max pooling) to define an information hierarchy. In practice, convolutional layers are also used to reduce the number of connections compared to fully connected layers, which can become impractically large for high-dimensional inputs.

Notably, some of the most popular object segmentation model~\cite{simonyan2014very, krizhevsky2012imagenet} and other breakthroughs such as  VGGNet~\cite{simonyan2014very}, ResNet~\cite{he2016deep} rely heavily on the power of CNNs. The seminal contributions of these latter underscore the capability of CNNs to extract meaningful structural information, motivating our exploration of convolution for mutual information estimation. 

By leveraging CNNs, this work aims to better capture localized structural patterns between plain and encrypted images in an attempt to gain insights to further comprehend information leakage in high dimensional data.

For this study, we focus on the usage of 
CLIP~\cite{radford2021learning} with the aim of assessing whether incorporating semantic information about the image content can improve leakage quantification within the MINE framework. The motivation behind this is to experiment with representations that capture not only spatial but also semantic context, with the hypothesis that such high-level understanding of image content might provide a more accurate estimation of the underlying information leakage. Prior work has demonstrated that CLIP captures high-level semantic representations of image content~\cite{goh2021multimodal, zhou2022learning}. 

We also evaluate the performance against a hierarchical visual feature extractor. We chose Resnet\cite{he2016deep} to assess the performance when relying solely on visual features and to compare with a different CNN architecture. 



Note that CNN embedding usually requires small input size and one would prefer to estimated the MI not on small resized version of images but on the original distributions. Therefore, we propose to sample the original distributions using image patches without any preprocessing such as resizing.

\section{Experiments}\label{sec : exp}

For our experiments, we have selected the 100 largest images (by resolution) from the COCO dataset~\cite{lin2014microsoft}. 
These images are encrypted using the method presented in Section~\ref{section:ise} with $s \in [0,8]$. Note that in our experiment, we can replace Eq.~\eqref{eq: sel_enc} by: $p_{\textsc{c}}  = p_{\textsc{e}} \ll s$. Indeed, this choice does not affect MI estimation but scales the clear part to larger values, which may be more natural for image embeddings.


The baseline experiment is the empirical estimation of the MI between an image and its encrypted version as a function of the number of encrypted bits $s$, and this result is illustrated in Fig.~\ref{sub:mipix}. We here recall that both MINE and the empirical method require large image size to produce significant results, since  increasing the size allows a better  approximation of the real pixel distribution.
We observe the linear behavior expected for random variable, described in Section~\ref{section:mc},  which confirms that empirical MI estimation on pixel histogram does not capture the specificity of images and it is clearly not correlated with  human perception, see Fig.~\ref{fig:selective-encryption}. 

\begin{figure}[h!]
    \begin{center}
    \subfloat[Cosine Similarity CLIP]{
        \includegraphics[width=0.5\columnwidth]{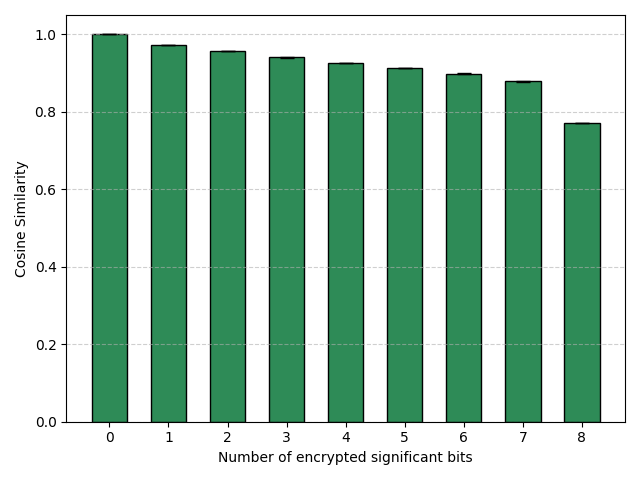}
        \label{sub:cossim}
    }
    \subfloat[MI CLIP]{
        \includegraphics[width=0.5\columnwidth]{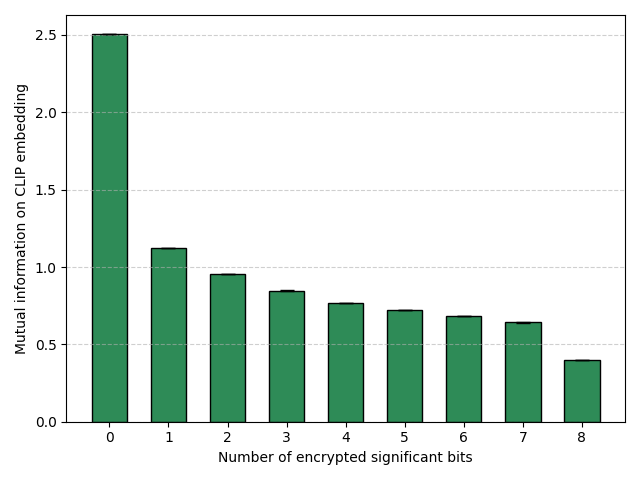}
        \label{sub:miclip}
    }
    \caption{On 100 images from the COCO dataset~\cite{lin2014microsoft}, average scores in terms of a) Cosine Similarity on the CLIP features, b) Mutual Information on the CLIP features, as a function of the number of encrypted bits $s$.}
    \label{fig:basic}
    \end{center}
\end{figure}

As a second step, we wanted to verify if using CNN may help to capture dependencies between pixels. Therefore, we used CLIP image embedding to compare an image and its encrypted counterpart in a most well-suited space. 
Note that images are resized to $224 \times 224$. This aligns with CLIP's initial training size and ensures consistent feature extraction to avoid any random internal downsampling \cite{radford2021learning}. 
Using a natural metric in this space, such as the cosine similarity, leads to results presented in Fig.~\ref{sub:cossim}. We conclude this metric is not interesting in our use case.

Nevertheless, we can estimate empirically the MI between the CLIP features, this is done by discretizing the features using naive rounding. Results, presented in Fig.~\ref{sub:miclip}, show that this method seems to capture more the non linearity of the expected MI estimation.

\begin{figure}[t]
    \centering
    \includegraphics[width=1\linewidth]{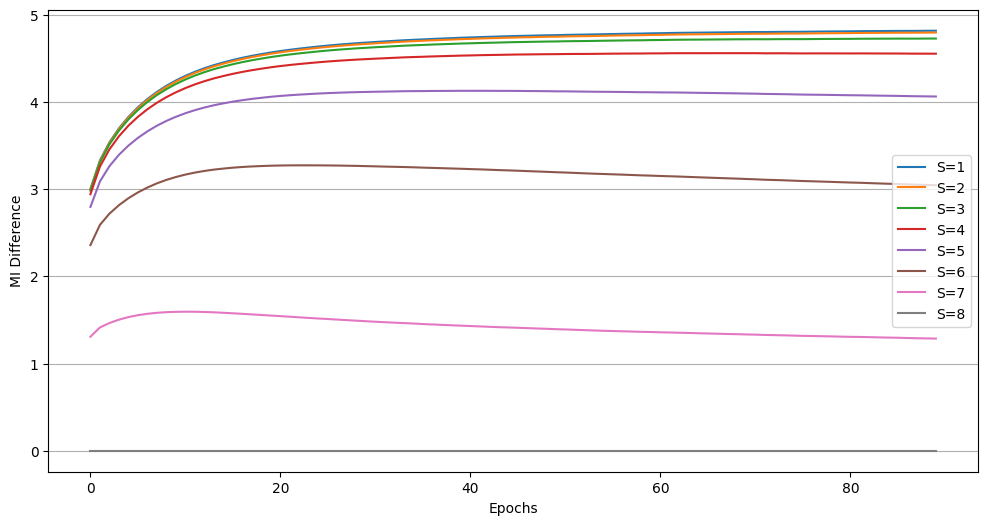}
    \caption{Relative distance to the maximum MI over 100 epochs on pixel distribution  using the MINE framework.}
    \label{fig:mi-diff-empirical}
\end{figure}

\begin{figure}[t]
    \centering
    \includegraphics[width=1\linewidth]{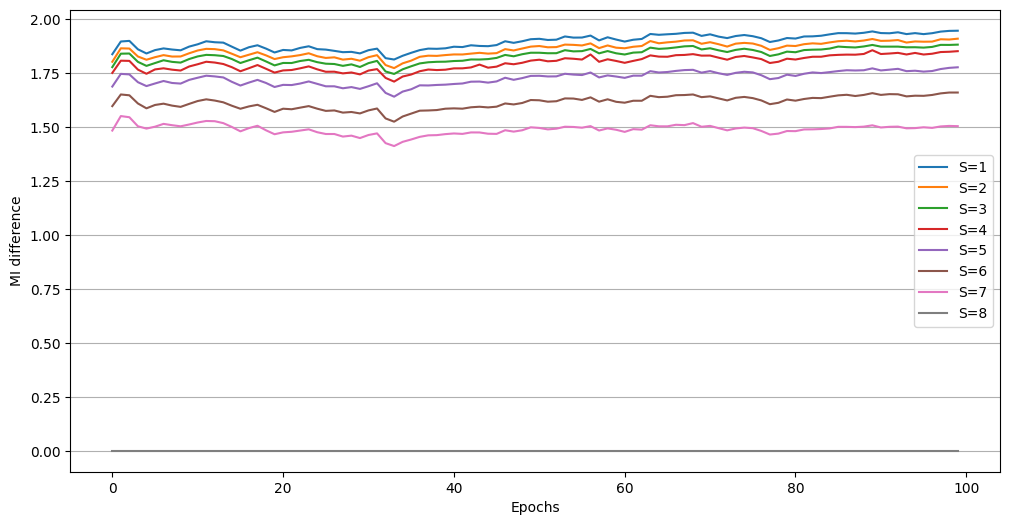}
    \caption{Relative distance to the maximum MI, computed via the MINE framework, over $100$ epochs using CLIP latent features. Subtracting the max MI highlights the behavior with respect to different $s$ values.}
    \label{fig:mi-diff-clip}
\end{figure}

\begin{figure}[t]
    \begin{center}
    \subfloat[Pixel distribution]{
        \includegraphics[width=0.5\columnwidth]{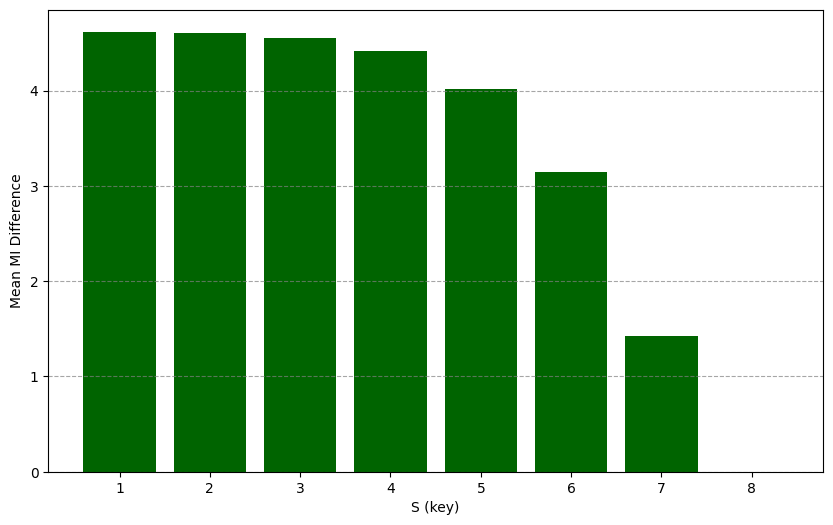}
        \label{sub:mine-max-mi}
    }
    \subfloat[CLIP latent features]{
        \includegraphics[width=0.5\columnwidth]{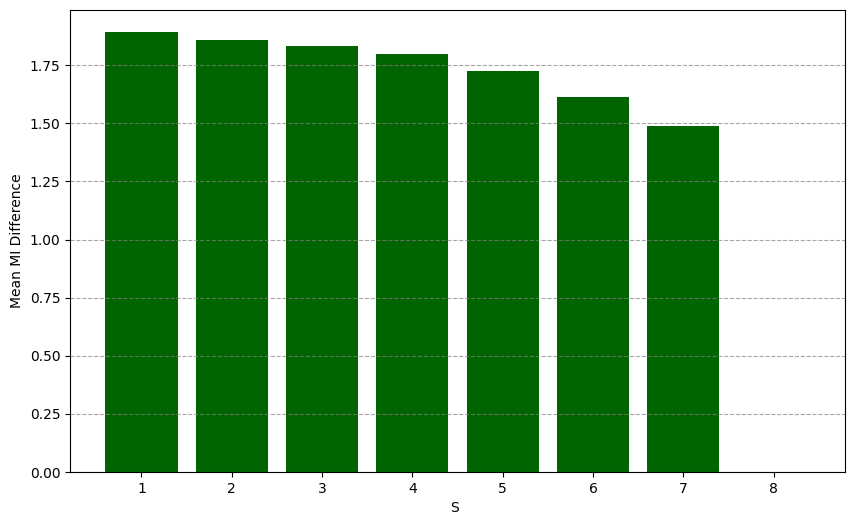}
        \label{sub:mi-clip-50}
    }
    \caption{On 100 images from the COCO dataset~\cite{lin2014microsoft}, MI computed with MINE framework on a) Pixel distribution and b) CLIP latent features for different $s$ levels.}
    \label{fig:mine-clip}
    \end{center}
\end{figure}

\begin{figure}[t]
    \centering
    \includegraphics[width=1\linewidth]{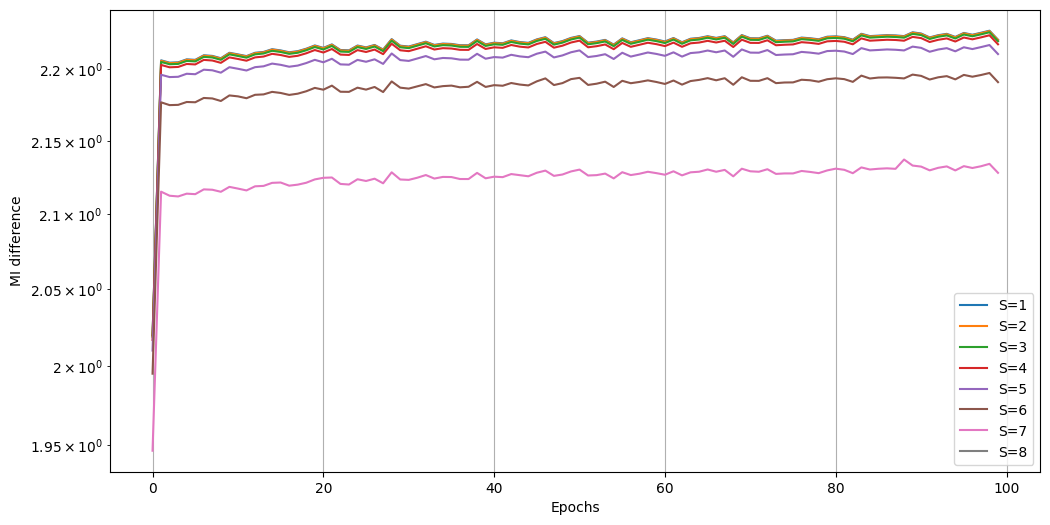}
    \caption{Relative distance to the maximum MI, on Resnet latent features via the MINE framework for different $s$ levels over $100$ epochs.}
    \label{fig:resnet-speed}
\end{figure}

\begin{figure}[t]
    \centering
    \includegraphics[width=1\linewidth]{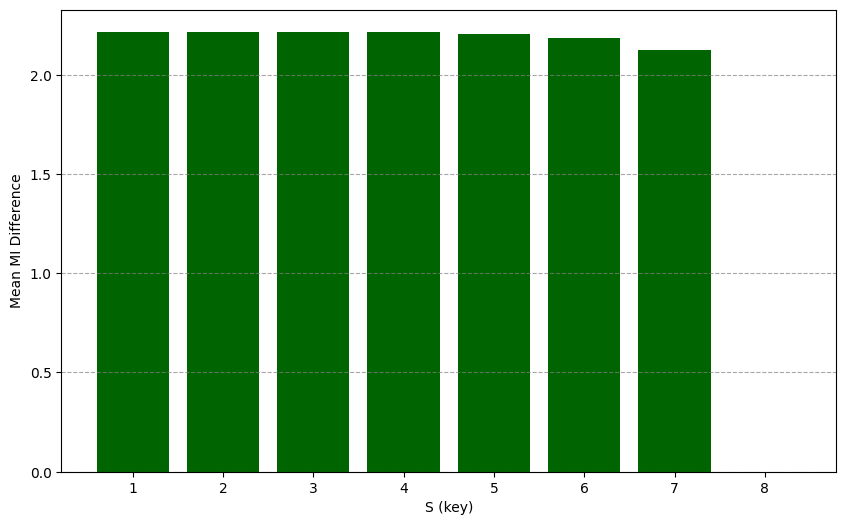}
    \caption{Relative distance to the maximum MI, computed via the MINE framework,  using Resnet latent features for different $s$ levels.}
    \label{fig:mi-diff-resnet}
\end{figure}

Now, in a new experiment we compare these results with the estimation of the MI using MINE. We estimate mutual information (MI) using MINE under two settings: \textit{(i)} directly on flattened pixel values and \textit{(ii)} on latent representations extracted via pretrained feature encoders CLIP ViT-B  and Resnet respectfully~\cite{radford2021learning, he2016deep}. In both cases, images were partially encrypted by masking a fixed number of bits $s$, and MI was tracked over $100$ epochs. MINE using CLIP,and Resnet features, is done by embedding patches of images. We have extracted an experimentally fixed $N=50$ number of patches of size $224 \times 224$. This choice aligns with the standard input size used during the pretraining of models such as CLIP ViT-B on ImageNet, where images are typically resized and center-cropped to $224 \times 224$~\cite{krizhevsky2012imagenet}. Both the pixel-level, CLIP-based and Resnet estimations were conducted over the same number of training epochs to ensure comparability. For each estimator, we plot the relative distance to the maximum MI — defined as the absolute difference from the highest value across all \( s \) levels — to better highlight subtle differences, particularly at lower \( s \) values where the curves are closely packed. This transformation allows a more detailed inspection of how intermediate encryption levels behave during training. Results are presented, respectively, in Fig.~\ref{fig:mi-diff-empirical},~\ref{fig:mi-diff-clip},~\ref{fig:mine-clip},~\ref{fig:resnet-speed} and~\ref{fig:mi-diff-resnet}.

First, as shown in Fig.~\ref{fig:mi-diff-empirical}, we observe that MINE’s estimation of mutual information between the original image and the clear portion of the encrypted one does not follow a linear trend and requires a large number of epochs to converge.

In contrast, when applied to CLIP latent representations (Fig.~\ref{fig:mi-diff-clip}), MINE demonstrates significantly faster convergence compared to the pixel-based setting.

While the reason behind the accelerated convergence on feature maps is less clear, it may be attributed to the inductive biases or pretraining of the encoder, which produces smoother or more separable representations, even in the presence of structured noise.

Perhaps more interestingly,  in Fig.~\ref{fig:mine-clip} we notice, for intermediate encryption levels  $s=6$ and $s=7$, higher MI estimations compared to the pixel distribution (Fig.~\ref{sub:mi-clip-50} and \ref{sub:mine-max-mi} respectively), hinting to CLIP's ability to capture the semantic underlying structure of neighboring pixels.
More to this point, we observe in Fig.~\ref{fig:mi-diff-clip} that for higher encryption levels, the MI curves remain more distinct and granular when using latent features. We hypothesize that this behavior stems from CLIP’s capacity to seize residual structure or patterns even in heavily perturbed inputs.

Additionally to CLIP, we have conducted some experiments with Resnet, in order to investigate other approaches. 
From Fig.~\ref{fig:resnet-speed} we can see that estimation on Resnet features shows an accelerated convergence compared to the initial pixel distribution. 
Furthermore, as depicted in  Fig.~\ref{fig:mi-diff-resnet},  we observe that Resnet, with slightly perturbed inputs, manages to capture structure on early $s$ levels, whereas it produces lower MI values as $s$ increases. 

Comparing the results of Fig.~\ref{fig:resnet-speed}  with the ones of Fig.~\ref{fig:mi-diff-clip}, we see that there is a different decrease in MI as $s$ changes, and the encrypted images loose structure.
We hypothesize that this difference between CLIP and Resnet's MI estimations could be caused by both CLIP's ability to capture semantics in the encoding, and Resnet's sensitivity to noise.

This leaves both encoders candidates for further investigation. 

\section{Conclusion and perspectives}

In this work, we have evaluated information leakage in selectively encrypted images using Mutual Information Neural Estimation (MINE), addressing structural limitations by incorporating convolutional processing to better capture spatial dependencies.

 Our investigation suggests that feature extractors may preserve structured signals even in partially encrypted data, therefore opening new opportunities for measuring information leakage in compressed or transformed domains.

However, this direction raises open questions regarding the quality, dimensionality of the latent representations and inclusivity of semantics. It also raises questions about noise sensitivity for visual feature encoders, which leaves the room to investigate visual feature encoders trained on partially noisy images. The effectiveness of MI estimation in this space also appears to be sensitive to several hyperparameters, including the choice of feature extractor, training schedule, and number of epochs. While this work lays the groundwork for latent-space MI analysis, we believe that further investigation — through careful tuning and ablation studies — could provide deeper insights into the robustness and limitations of this approach. 

Moreover, while this work represents an initial step from the cryptographic perspective, a compelling and challenging direction for future research lies in leveraging encryption paradigms themselves to further inform and enhance mutual information-based leakage analysis.


\section{Acknowledgment}
This work was funded by the French National Research Agency (ANR), under contract ANR-23-IAS4-0004 project CI2(IA).

\bibliographystyle{IEEEtran}
\bibliography{references}

\begin{thebibliography}{10}
\providecommand{\url}[1]{#1}
\csname url@samestyle\endcsname
\providecommand{\newblock}{\relax}
\providecommand{\bibinfo}[2]{#2}
\providecommand{\BIBentrySTDinterwordspacing}{\spaceskip=0pt\relax}
\providecommand{\BIBentryALTinterwordstretchfactor}{4}
\providecommand{\BIBentryALTinterwordspacing}{\spaceskip=\fontdimen2\font plus
\BIBentryALTinterwordstretchfactor\fontdimen3\font minus
  \fontdimen4\font\relax}
\providecommand{\BIBforeignlanguage}[2]{{%
\expandafter\ifx\csname l@#1\endcsname\relax
\typeout{** WARNING: IEEEtran.bst: No hyphenation pattern has been}%
\typeout{** loaded for the language `#1'. Using the pattern for}%
\typeout{** the default language instead.}%
\else
\language=\csname l@#1\endcsname
\fi
#2}}
\providecommand{\BIBdecl}{\relax}
\BIBdecl

\bibitem{liu2020privacy}
X.~Liu, L.~Xie, Y.~Wang, J.~Zou, J.~Xiong, Z.~Ying, and A.~V. Vasilakos,
  ``Privacy and security issues in deep learning: A survey,'' \emph{IEEE
  Access}, vol.~9, pp. 4566--4593, 2020.

\bibitem{huang2022privacy}
Q.-X. Huang, W.~L. Yap, M.-Y. Chiu, and H.-M. Sun, ``Privacy-preserving deep
  learning with learnable image encryption on medical images,'' \emph{IEEE
  Access}, vol.~10, pp. 66\,345--66\,355, 2022.

\bibitem{ahmad2010efficiency}
J.~Ahmad and F.~Ahmed, ``Efficiency analysis and security evaluation of image
  encryption schemes,'' \emph{Computing}, vol.~23, no.~4, p.~25, 2010.

\bibitem{dwork2006differential}
C.~Dwork, ``Differential privacy,'' in \emph{International colloquium on
  automata, languages, and programming}.\hskip 1em plus 0.5em minus 0.4em\relax
  Springer, 2006, pp. 1--12.

\bibitem{dwork2008differential}
------, ``Differential privacy: A survey of results,'' in \emph{International
  conference on theory and applications of models of computation}.\hskip 1em
  plus 0.5em minus 0.4em\relax Springer, 2008, pp. 1--19.

\bibitem{dwork2014algorithmic}
C.~Dwork and A.~Roth, ``The algorithmic foundations of differential privacy,''
  \emph{Foundations and trends{\textregistered} in theoretical computer
  science}, vol.~9, no. 3--4, pp. 211--407, 2014.

\bibitem{bloch_overview_2021}
M.~Bloch, O.~G{\"u}nl{\"u}, A.~Yener, F.~Oggier, H.~V. Poor, L.~Sankar, and
  R.~F. Schaefer, ``An overview of information-theoretic security and privacy:
  Metrics, limits and applications,'' \emph{IEEE Journal on Selected Areas in
  Information Theory}, vol.~2, no.~1, pp. 5--22, 2021.

\bibitem{wang2016relation}
W.~Wang, L.~Ying, and J.~Zhang, ``On the relation between identifiability,
  differential privacy, and mutual-information privacy,'' \emph{IEEE
  Transactions on Information Theory}, vol.~62, no.~9, pp. 5018--5029, 2016.

\bibitem{kim2024crypto}
B.~D. Kim, V.~A. Vasudevan, J.~Woo, A.~Cohen, R.~G. D’Oliveira, T.~Stahlbuhk,
  and M.~M{\'e}dard, ``Crypto-mine: Cryptanalysis via mutual information neural
  estimation,'' in \emph{2024 IEEE International Conference on Acoustics,
  Speech and Signal Processing (ICASSP)}.\hskip 1em plus 0.5em minus
  0.4em\relax IEEE, 2024, pp. 4820--4824.

\bibitem{puteaux2021combining}
P.~Puteaux, V.~Itier, and P.~Bas, ``Combining forensics and privacy
  requirements for digital images,'' in \emph{2021 29th European Signal
  Processing Conference (EUSIPCO)}.\hskip 1em plus 0.5em minus 0.4em\relax
  IEEE, 2021, pp. 806--810.

\bibitem{larkin2016reflections}
K.~G. Larkin, ``Reflections on shannon information: In search of a natural
  information-entropy for images,'' \emph{arXiv preprint arXiv:1609.01117},
  2016.

\bibitem{belghazi2018mine}
M.~I. Belghazi, A.~Baratin, S.~Rajeshwar, S.~Ozair, Y.~Bengio, A.~Courville,
  and D.~Hjelm, ``Mutual information neural estimation,'' in
  \emph{International conference on machine learning}.\hskip 1em plus 0.5em
  minus 0.4em\relax PMLR, 2018, pp. 531--540.

\bibitem{sun2022survey}
R.~Sun, T.~Lei, Q.~Chen, Z.~Wang, X.~Du, W.~Zhao, and A.~K. Nandi, ``Survey of
  image edge detection,'' \emph{Frontiers in Signal Processing}, vol.~2, p.
  826967, 2022.

\bibitem{ding2024edge}
J.~Ding, J.-C. Zhao, Y.-Z. Sun, P.~Tan, J.-W. Wang, J.-E. Ma, and Y.~Fang,
  ``Edge detectors can make deep convolutional neural networks more robust,''
  \emph{Available at SSRN 4734191}.

\bibitem{zhang2018unreasonable}
R.~Zhang, P.~Isola, A.~A. Efros, E.~Shechtman, and O.~Wang, ``The unreasonable
  effectiveness of deep features as a perceptual metric,'' in \emph{2018 IEEE
  conference on computer vision and pattern recognition}.\hskip 1em plus 0.5em
  minus 0.4em\relax IEEE, 2018, pp. 586--595.

\bibitem{lookabaugh2004selective}
T.~Lookabaugh, ``Selective encryption, information theory and compression,'' in
  \emph{Conference Record of the Thirty-Eighth Asilomar Conference on Signals,
  Systems and Computers}, vol.~1.\hskip 1em plus 0.5em minus 0.4em\relax IEEE,
  2004, pp. 373--376.

\bibitem{lin2014microsoft}
T.-Y. Lin, M.~Maire, S.~Belongie, J.~Hays, P.~Perona, D.~Ramanan,
  P.~Doll{\'a}r, and C.~L. Zitnick, ``Microsoft coco: Common objects in
  context,'' in \emph{13th European conference on computer vision}.\hskip 1em
  plus 0.5em minus 0.4em\relax Springer, 2014, pp. 740--755.

\bibitem{younesi2024comprehensive}
A.~Younesi, M.~Ansari, M.~Fazli, A.~Ejlali, M.~Shafique, and J.~Henkel, ``A
  comprehensive survey of convolutions in deep learning: Applications,
  challenges, and future trends,'' \emph{IEEE Access}, vol.~12, pp.
  41\,180--41\,218, 2024.

\bibitem{deneu2021convolutional}
B.~Deneu, M.~Servajean, P.~Bonnet, C.~Botella, F.~Munoz, and A.~Joly,
  ``Convolutional neural networks improve species distribution modelling by
  capturing the spatial structure of the environment,'' \emph{PLoS
  computational biology}, vol.~17, no.~4, 2021.

\bibitem{krizhevsky2012imagenet}
A.~Krizhevsky, I.~Sutskever, and G.~E. Hinton, ``Imagenet classification with
  deep convolutional neural networks,'' \emph{Advances in neural information
  processing systems}, vol.~25, 2012.

\bibitem{simonyan2014very}
K.~Simonyan and A.~Zisserman, ``Very deep convolutional networks for
  large-scale image recognition,'' \emph{arXiv preprint arXiv:1409.1556}, 2014.

\bibitem{he2016deep}
K.~He, X.~Zhang, S.~Ren, and J.~Sun, ``Deep residual learning for image
  recognition,'' in \emph{2016 IEEE conference on computer vision and pattern
  recognition}.\hskip 1em plus 0.5em minus 0.4em\relax IEEE, 2016, pp.
  770--778.

\bibitem{radford2021learning}
A.~Radford, J.~W. Kim, C.~Hallacy, A.~Ramesh, G.~Goh, S.~Agarwal, G.~Sastry,
  A.~Askell, P.~Mishkin, J.~Clark \emph{et~al.}, ``Learning transferable visual
  models from natural language supervision,'' in \emph{International conference
  on machine learning}.\hskip 1em plus 0.5em minus 0.4em\relax PMLR, 2021, pp.
  8748--8763.

\bibitem{goh2021multimodal}
G.~Goh, N.~Cammarata, C.~Voss, S.~Carter, M.~Petrov, L.~Schubert, A.~Radford,
  and C.~Olah, ``Multimodal neurons in artificial neural networks,''
  \emph{Distill}, vol.~6, no.~3, 2021.

\bibitem{zhou2022learning}
K.~Zhou, J.~Yang, C.~C. Loy, and Z.~Liu, ``Learning to prompt for
  vision-language models,'' \emph{International journal of computer vision},
  vol. 130, no.~9, pp. 2337--2348, 2022.

\end{thebibliography}
\end{document}